\documentclass{jaa}

\usepackage{graphicx}
\usepackage{xspace}
\usepackage{epsf,url,wrapfig ,fancyhdr,multirow,lscape,appendix,rotating, setspace,amsmath,amssymb}
\usepackage{graphics,epsfig,url}
\usepackage{graphicx}
\usepackage{gensymb}
\usepackage{amssymb}
\usepackage{xcolor}
\usepackage{color}
\usepackage{lipsum}
\usepackage{textcomp}

\newcommand{\msol}{\ensuremath{M_{\odot}}\xspace}

\newcommand{\rxte}{\textit{RXTE}\xspace}
\newcommand{\nustar}{\textit{NuSTAR}\xspace}
\newcommand{\astrosat}{\textit{AstroSat}\xspace}

\newcommand{\swift}{\textit{Swift}\xspace}

\newcommand{\source}{EXO~2030+375~}
\newcommand{\ms}{$M_{\odot}$}

\newcommand{\fluxcgs}{erg~s$^{-1}$~cm$^{-2}$}
\newcommand{\lumcgs}{erg~s$^{-1}$}

\begin{document}\sloppy


\title{Detection of X-ray pulsations at the lowest observed luminosity of Be/X-ray binary pulsar EXO~2030+375 with {\em AstroSat}}

\author{Gaurava K. Jaisawal\textsuperscript{1}, Sachindra Naik\textsuperscript{2}, Shivangi Gupta\textsuperscript{2}, P. C. Agrawal\textsuperscript{3}, Arghajit Jana\textsuperscript{2}, Birendra Chhotaray\textsuperscript{2,4} and Prahlad R. Epili\textsuperscript{5}}
\affilOne{\textsuperscript{1}National Space Institute, Technical University of Denmark, Elektrovej 327-328, DK-2800 Lyngby, Denmark\\}
\affilTwo{\textsuperscript{2}Astronomy and Astrophysics Division, Physical Research Laboratory, Navrangpura, Ahmedabad - 380009, India\\}
\affilThree{\textsuperscript{3}Department of Astronomy and Astrophysics (Retired), Tata Institute of Fundamental Research, Colaba, Mumbai - 400005, India\\}
\affilFour{\textsuperscript{4}Indian Institute of Technology, Gandhinagar, Gujarat, India\\}
\affilFive{\textsuperscript{5}School of Physics and Technology, Wuhan University, Wuhan 430072, China\\}

\twocolumn[{

\maketitle

\corres{gaurava@space.dtu.dk}

\msinfo{***}{***}

\begin{abstract}
We present the results obtained from timing and spectral studies of Be/X-ray binary pulsar EXO~2030+375 using observations with the Large Area Xenon Proportional Counters and Soft X-ray Telescope of {\em AstroSat}, at various phases of its Type-I outbursts in 2016, 2018, and 2020. The pulsar was faint during these observations as compared to earlier observations with other observatories. At the lowest luminosity of 2.5$\times$10$^{35}$ erg s$^{-1}$ in 0.5--30 keV energy range, $\approx$41.3~s pulsations were clearly detected in the X-ray light curves. This finding establishes the first firm detection of pulsations in EXO~2030+375 at an extremely low mass accretion rate to date. The shape of the pulse profiles is complex due to the presence of several narrow dips. Though pulsations were detected up to $\sim$80 keV when the source was brighter, pulsations were limited up to $\sim$25 keV during the third {\em AstroSat} observation at lowest source luminosity. A search for quasi-periodic oscillations in 2$\times$10$^{-4}$ Hz to 10 Hz yielded a negative result. Spectral analysis of the \astrosat data showed that the spectrum of the pulsar was steep with a power-law index of $\sim$2. The values of photon-indices at observed low luminosities follow the known pattern in sub-critical regime of the pulsar.

\end{abstract}


}]


\doinum{12.3456/s78910-011-012-3}
\artcitid{\#\#\#\#}
\volnum{000}
\year{0000}
\pgrange{1--}
\setcounter{page}{1}
\lp{1}

\section{Introduction}

Accreting high mass X-ray binary (HMXB) pulsars are among the brightest X-ray sources in our Galaxy (Nagase et al. 1989). In these binaries, a neutron star and a massive ($\geq$ 10 \ms) main-sequence star rotate around the common center of mass of the system in a wide and eccentric orbit (Tauris \& van den Heuvel 2006). The neutron star accretes matter from the companion star through the capture of stellar wind or Roche-lobe overflow. A majority of the HMXB systems are known to be Be/X-ray binaries (BeXRBs) in which the mass-donor is a non-supergiant B or O spectral type star which shows emission lines in its optical/infrared spectrum (Reig 2011). Rapid rotation of the companion Be star in the BeXRB system expels its photospheric matter equatorially, forming a circumstellar disk around it. The continuously evolving, equatorial circumstellar disk is known to be the cause of the emission lines and infrared excess in the optical/infrared spectrum of the companion star in the BeXRBs. Significant evolution of the circumstellar disk allows the neutron star to capture copious amount of matter while passing through the periastron. This abrupt accretion of matter by the neutron star enhances its X-ray emission by several orders of magnitude which lasts for several days to tens of days. These events are termed as Type-I X-ray outbursts. Once the neutron star moves away from the periastron, accretion from the circumstellar disk is no more possible and the X-ray source returns to the quiescence. The long term X-ray activity in BeXRBs is characterized by the regular Type-I outbursts with peak luminosity of the order of $ L_{x} \le 10^{37}$~erg~s$^{-1}$ and irregular rare giant (Type-II) X-ray outbursts with peak luminosity of $L_{x} \geq 10^{37}$~erg~s$^{-1}$. The Type-I X-ray outbursts are of short duration, covering 20--30 \% of orbit and coincide with the periastron passage of the neutron star whereas the Type-II outbursts show no preferred orbital phase dependence but once set in, they tend to cover a large fraction of the orbital period or even several orbital periods (see, e.g., Okazaki \& Negueruela 2001, Reig 2011, Jaisawal \& Naik 2016, Wilson-Hodge et al. 2018, Jaisawal et al. 2019).

EXO~2030+375 is one of the well studied Be/X-ray binary pulsars associated with regular Type-I outbursts during almost every periastron passage. This transient accreting X-ray pulsar was discovered in 1985 with {\it EXOSAT} during a giant outburst (Parmar et. al. 1989) with $\sim$42~s pulsations. The transient behaviour of this pulsar could be traced since its discovery when its initial 1-20 keV outburst luminosity (1.0$\times$10$^{38}~d_{5}^{2}$)~erg~s$^{-1}$ on 1985 May 18 declined by a factor of $\ge$2600 within 100 days of the outburst. The associated optical counterpart of EXO~2030+375 is a  highly reddened B0 Ve star (Motch \& Janot-Pascheco 1987) showing infrared excess and H$\alpha$ in emission (Coe et al. 1988). Using the relationship between extinction and distance of sources in the Galactic plane, Wilson et al. (2002) estimated the distance of \source to be 7.1 kpc. The regular Type-I X-ray outbursts of EXO~2030+375, occurring almost at every periastron passage of its $\sim$46 day orbit (Wilson et al. 2008), have been extensively monitored with the X-ray instruments onboard {\it RXTE}, {\it INTEGRAL}, {\it XMM-Newton}, {\it Suzaku} and {\it Swift/BAT} observatories to understand the characteristic properties of the pulsar (Wilson et al. 2002; Naik et al. 2013; Naik \& Jaisawal 2015, Ferrigno et al. 2016;  Epili et al. 2017 and references therein).

In June 2006, EXO 2030+375 was caught for the second time in a giant (Type-II) X-ray   outburst with initial flux of 180 mCrab. This surpassed  the previous peak flux of about 50~mCrab observed during the entire life of the  {\em RXTE}/ASM mission (Corbet \& Levine 2006).
The 2006 Type-II outburst was also followed by {\em Swift}/BAT which reported the peak flux steadily increased to 750 mCrab (Krimm et. al. 2006). Spin-up trend was observed in the pulsar during the giant X-ray outbursts in 1985  (Parmar et al. 1989) and 2006 (Wilson, Finger \& Camero-Arranz 2008) whereas spin-down episodes have been observed at low luminous outbursts in 1994-2002 (Wilson et al. 2002; Wilson, Fabregatet \& Coburn 2005) and during faint outbursts after March 2016 (Kretschmar et al. 2016). 

The phase-averaged spectra of EXO~2030+375 during normal and giant outbursts prior to 2006 giant outburst were described with various phenomenological and physical models (in some cases) along with an iron emission line at 6.4~keV and interstellar absorption (Epili et al. 2017 and references therein). Apart from the continuum spectrum, several interesting features have also been observed in the pulsar spectrum. {\it Suzaku} observations of the pulsar EXO~2030+375 during less intense Type-I outbursts in 2007 and 2012 did not show any evidence of cyclotron absorption features in the X-ray spectrum of EXO~2030+375. However, presence of additional matter locked at certain pulse phases of the pulsar was reported and interpreted as the cause of several prominent absorption dips in the pulse profiles (Naik et al. 2013; Naik \& Jaisawal 2015). During the brighter Type-I outburst in 2007, Naik et al. (2013) detected several narrow emission lines (i.e Si~{\sc XII}, Si~{\sc XIV}, S~{\sc XV}) for the
first time along with Fe~K$_{\alpha}$ and Fe~{\sc XVI} in the X-ray spectrum.

\begin{table*}[bt!]
\tabularfont
\centering
\caption{Log of observations of \source with {\em AstroSat}, \nustar and \swift/XRT.}
\begin{tabular}{ccllcclc}
\hline
\hline
         &Observation ID           &\multicolumn{2}{c}{Start of Observation}      &\multicolumn{2}{c}{Exposure (in ks)}  &Spin period   &Count \\
         &                         &Date          &MJD    &LAXPC   &SXT     & (s)  &Rate $^b$\\
\hline
{\underline {\em AstroSat}}          \\
Obs-1   &G06\_089T01\_9000000746    &23 October 2016     &57684.99      &48.5  &12.1   &41.2895(7)   &32\\     
Obs-2    &G08\_081T01\_9000002144   &6 June 2018         &58275.53      &43.6  &24.1   &41.272(9)     &24\\
Obs-3   &G08\_081T01\_9000002178    &19 June 2018        &58288.88      &46.4  &23     &41.30(1)      &11\\
Obs-4   &G08\_081T01\_9000002350    &9 September 2018    &58370.3       &46.5  &22.8    &41.2747(8)    &65\\
Obs-5   &T03\_244T01\_9000003912    &2 October 2020      &59124.57      &95    &8.9     &41.306(3)     &27\\
\hline
\nustar  &90201029002               &25 July 2016        &57594.36   &\multicolumn{2}{c}{56.7}    &41.287054$^a$   &---\\
\swift  &00030799022                &25 July 2016        &57594.85   &\multicolumn{2}{c}{1}       &---    &---\\
 \hline
\hline
\end{tabular}
\label{log}
\tablenotes{$^a$: from F{\"u}rst et al. (2017). $^b$: Average source count rate (in counts s$^{-1}$) per LAXPC unit is given in 3-80 keV energy range. }
\end{table*}

\begin{figure}[bt!]
\centering
\includegraphics[width=0.5\textwidth, angle=0]{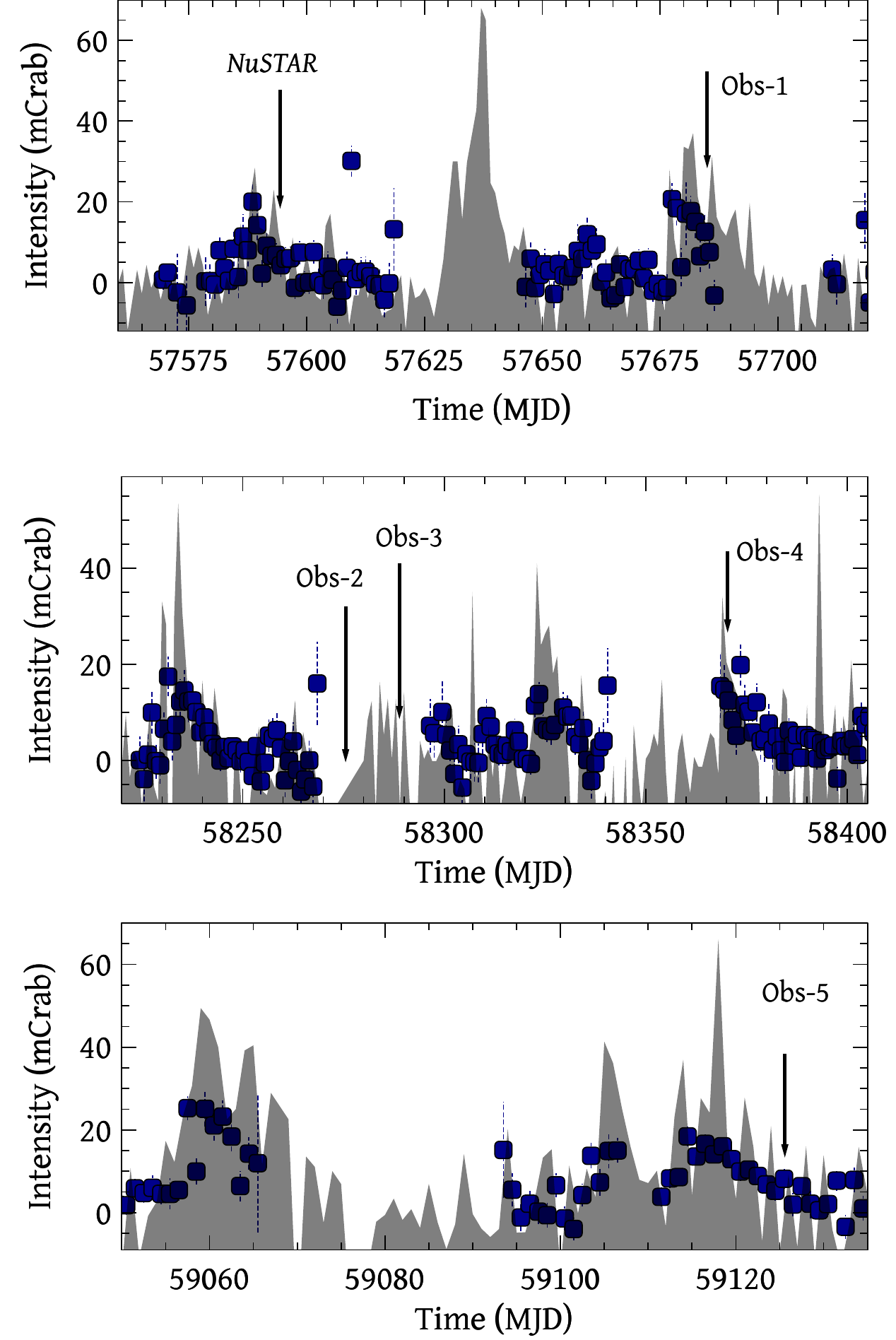}
\caption{MAXI (2-20 keV, blue data points) and \swift/BAT (15-50 keV, shaded) long-term monitoring light curves of \source ranging from (a). 21 June 2016 (MJD 57560) to 27 November 2016 (MJD 57719), (b). 12 April 2018 (MJD 58220) to 14 October 2018 (MJD 58405), and (c). 20 July 2020 (MJD 59050) to 13 October 2020 (MJD 59135) are shown in top, middle, and bottom panels, respectively. Arrow marks in the panels represent the epochs of \astrosat and \nustar observations of the pulsar. }
\label{maxi-bat}
\end{figure}

A detailed and comprehensive study of EXO~2030+375 was carried out by using extensive {\it RXTE} pointed observations during many Type-I and 2006 Type-II outbursts starting from 1995 till 2011 (Epili et al. 2017). Timing and spectral studies of the pulsar were carried out in 3--30 keV luminosity range from 3.8$\times$10$^{36}$ to 2.6$\times$10$^{38}$ erg s$^{-1}$, covered during the entire duration of {\it RXTE} campaign. Timing studies of more than 600 {\it RXTE} pointings revealed the evolution of pulse profiles of the pulsar with luminosity - a main peak and a minor peak at low luminosity evolved into a two-peaked profile along with minor dips at high luminosity. This study revealed that pulse profiles of the pulsar at a particular luminosity were identical irrespective of the type of X-ray outbursts, indicating that the emission geometry depends mainly on the accretion rate. 

Since the discovery in 1985, the pulsar had been showing regular X-ray outbursts for about 25 years. Since early 2015, however, the Type-I outbursts appeared to be of decreasing intensity and eventually vanished from the light curve towards the end of 2015 or early 2016 (F{\"u}rst  et al. 2016). The Type-I X-ray outburst activity commenced again in early 2016 and still continuing, though with much fainter peak luminosity ($\le$10$^{36}$ erg s$^{-1}$) than the usual ones. F{\"u}rst  et al. (2017) reported the detection of pulsation at a minimum luminosity of 6.8$\times$10$^{35}$ erg s$^{-1}$ in 3-78 keV range, considered to be the lowest luminosity of the pulsar with X-ray pulsations in the light curve. Though the pulsar was observed with \swift/XRT at a fainter phase, the data quality was not good enough for pulsation search. As the pulsar is still showing Type-I X-ray outbursts with fainter peak luminosity, it is interesting to carry out timing and spectral studies with {\it Astrosat}~  to explore whether the pulsar has gone into the propeller regime or still undergoing accretion. Detection of pulsations in the light curve at lower luminosity compared to that during the earlier \swift/XRT observation (F{\"u}rst  et al. 2017), would rule out the onset of propeller regime. Further, detection of pulsation at a limiting luminosity may allow us in estimating the magnetic field of the pulsar. The \astrosat observations at lower luminosity, therefore, are  important to investigate above properties of the pulsar. In this paper, we investigate the pulsation activities, shape of pulse profiles and spectral properties of the pulsar at a significantly lower luminosity level using five epochs of \astrosat observations. For comparison, data from \nustar observation of the pulsar on 25 July 2016, reported in F{\"u}rst  et al. (2017),  were also used in present work. The observations of the pulsar and data reduction procedures are described in Section~2, results obtained from the timing and spectral analysis are presented in Section~3 \& 4, respectively. The implication of our results are discussed in Section~5.

\begin{figure}[t!]
\centering
\includegraphics[height=2.6in, width=3.2in, angle=0]{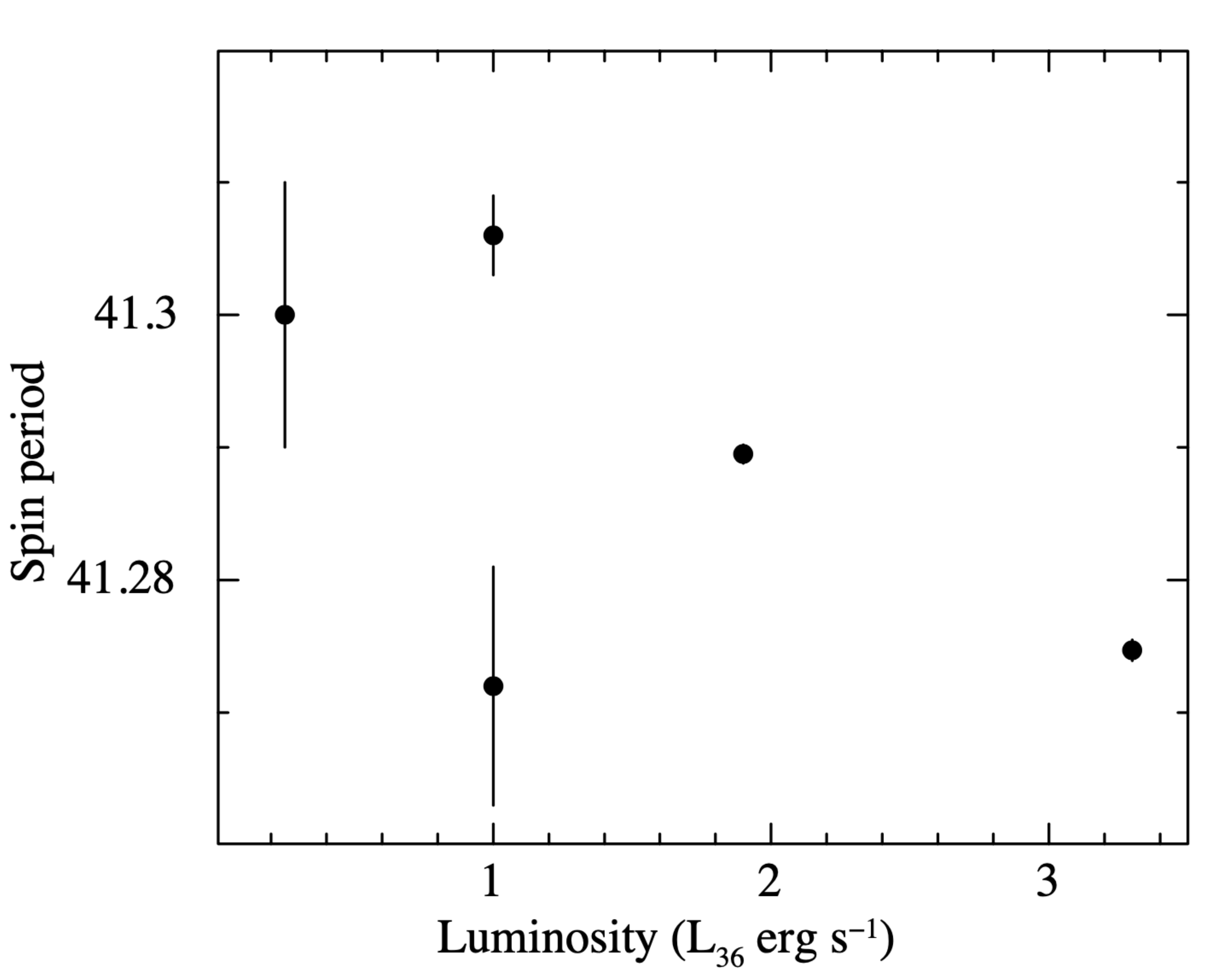}
\caption{Spin period evolution of the pulsar with luminosity during \astrosat observations. L$_{36}$ represents the 3-30 keV unabsorbed luminosity in unit of 10$^{36}$~\lumcgs.}
\label{spin-period}
\end{figure}

\begin{figure*}
\centering
\includegraphics[width=0.23\textwidth,angle=-90]{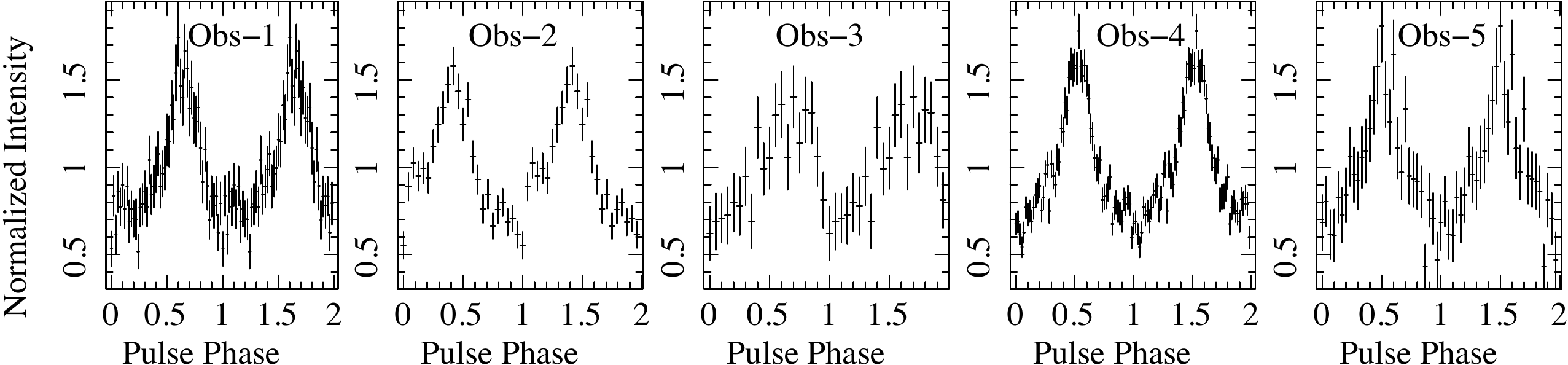}
\caption{ Pulse profiles of \source in 0.3-7 keV range with SXT instrument are shown for all five \astrosat observations (left  to right). These profiles are obtained by folding the light curves at respective pulse period determined from LAXPC data. Two pulses are shown for clarity.}
\label{sxt-profile}
\end{figure*}

\begin{figure}[bt!]
\centering
\includegraphics[height=4.8in, width=3.2in, angle=0]{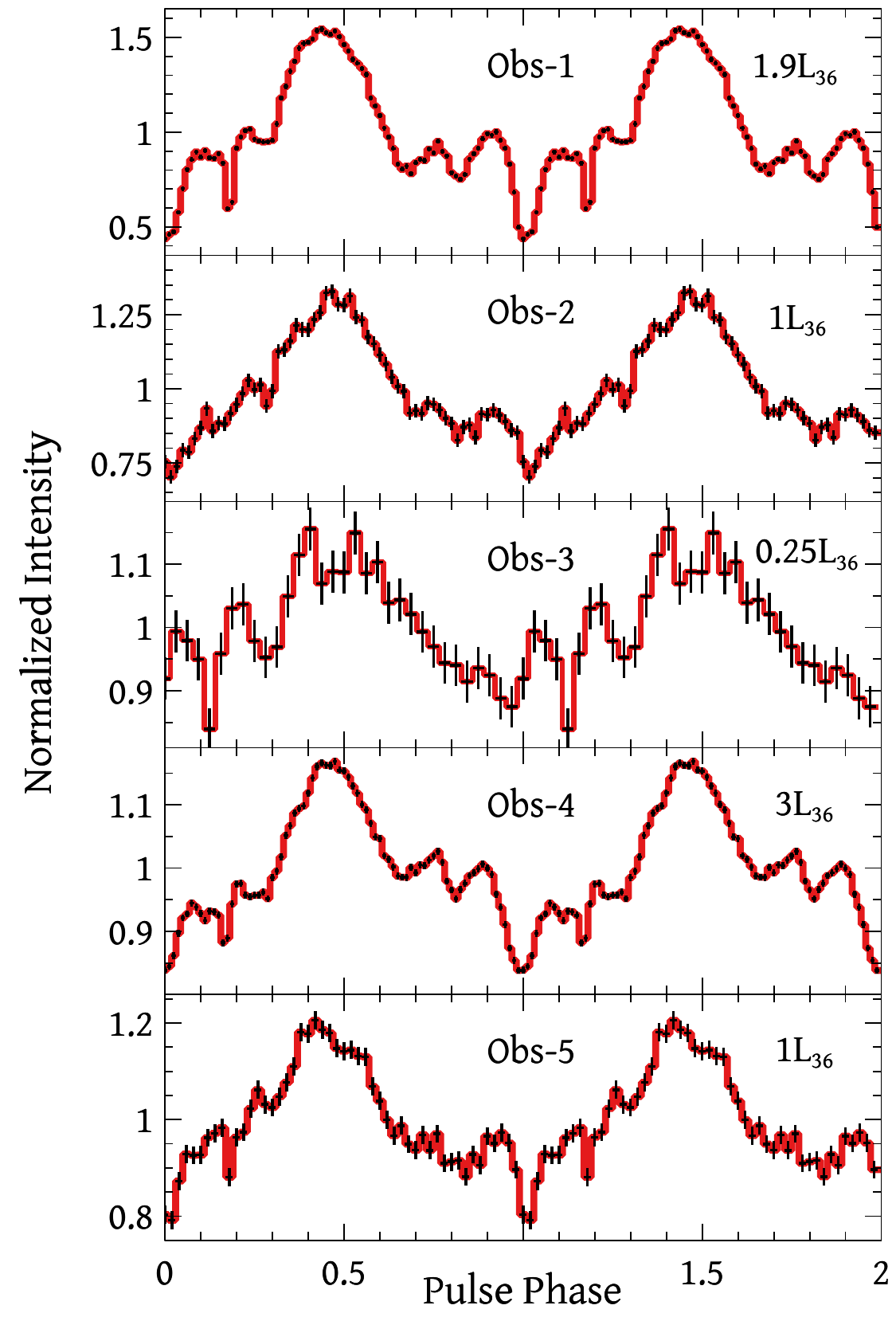}
\caption{ Pulse profiles of \source in 3-80 keV range are shown for all five \astrosat observations (top to bottom). L$_{\rm 36}$ denotes the 0.5-30 keV unabsorbed luminosity of the pulsar in 10$^{\rm 36}$~\lumcgs at a distance of 7.1~kpc. Two pulses are shown for clarity.}
\label{lxp-profile}
\end{figure}

\begin{figure*}[bt!]
\centering
\includegraphics[width=.5\textwidth,angle=-90]{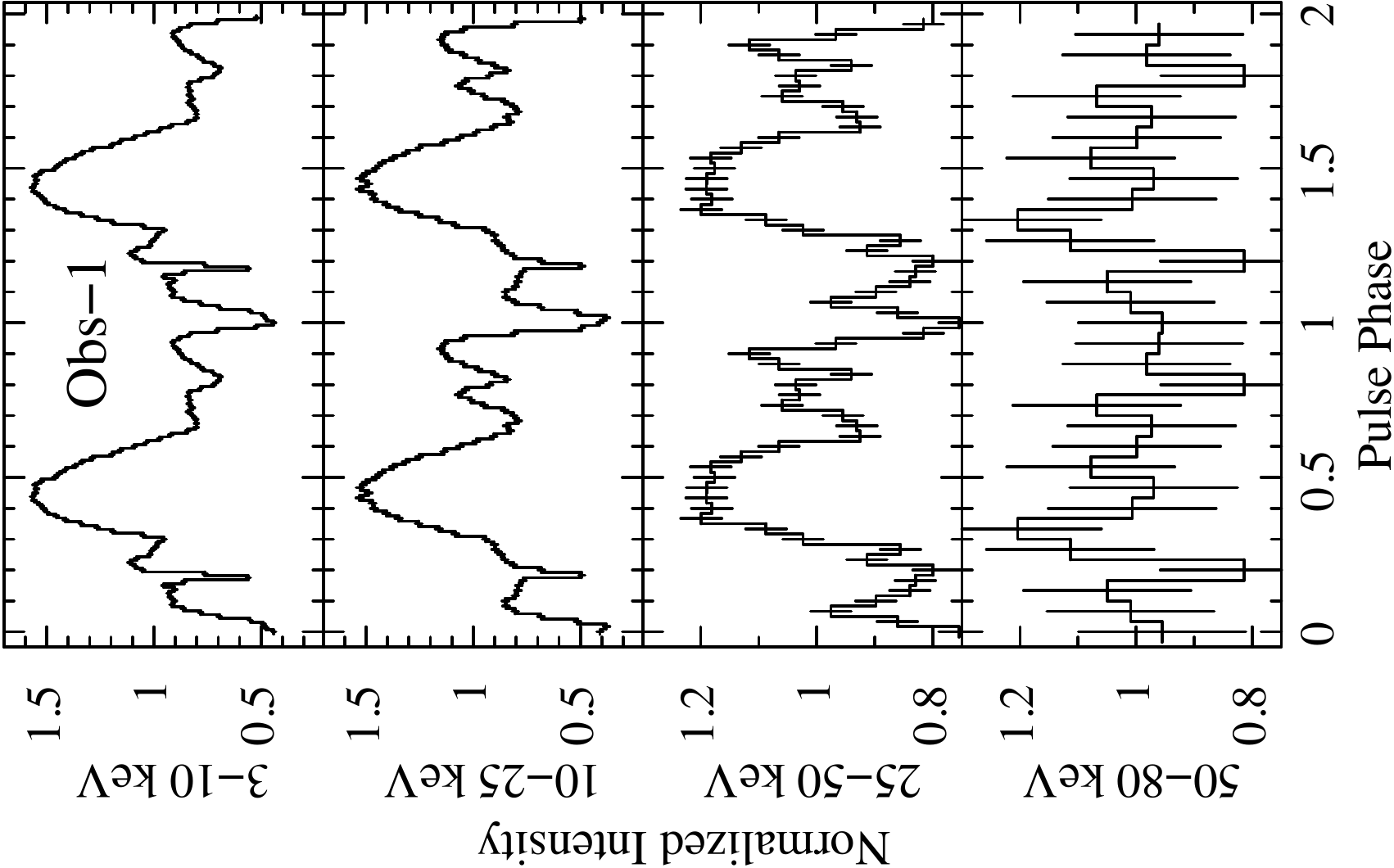}\quad
\includegraphics[width=.5\textwidth, angle=-90]{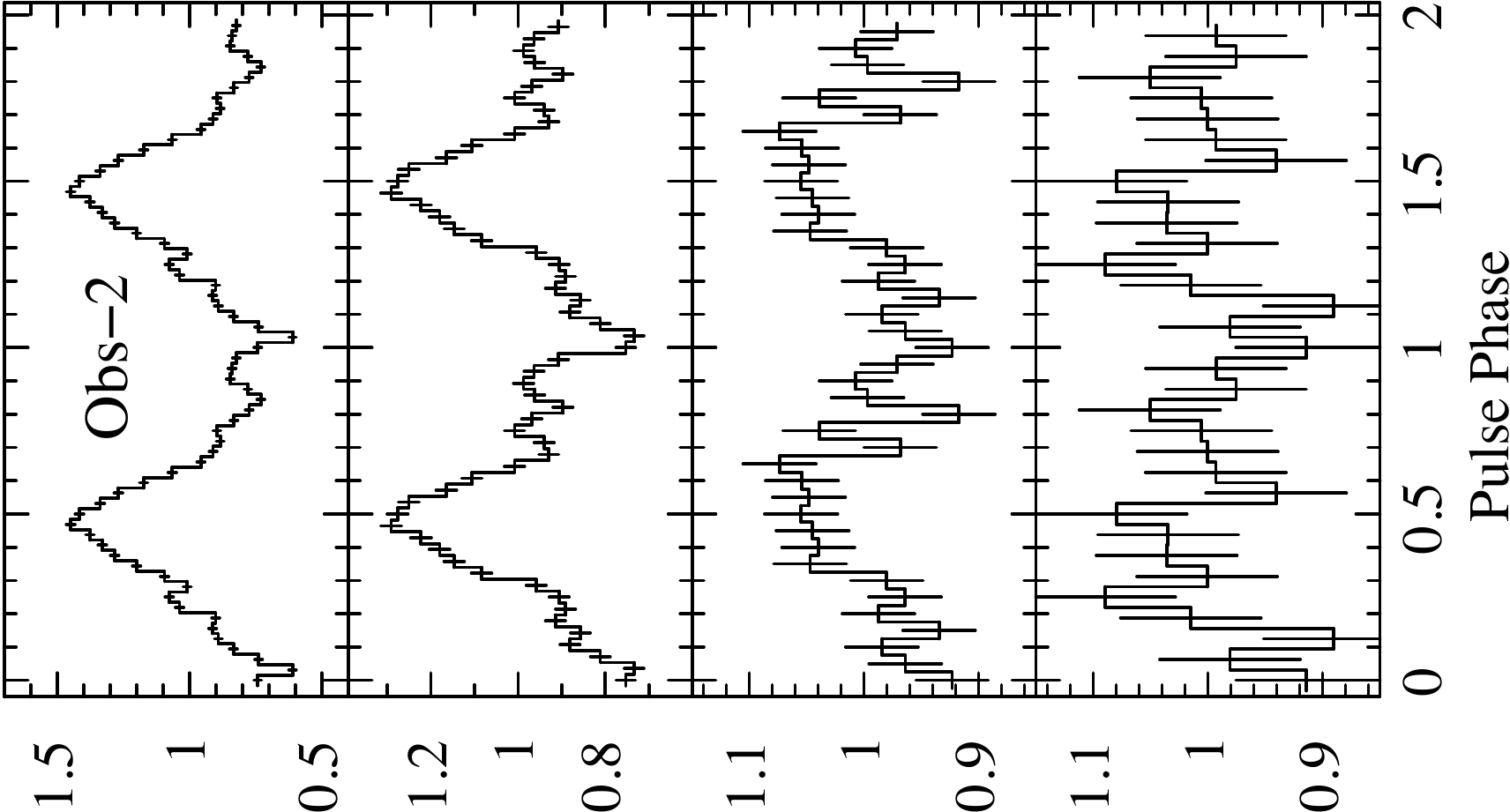}\quad
\includegraphics[width=.5\textwidth, angle=-90]{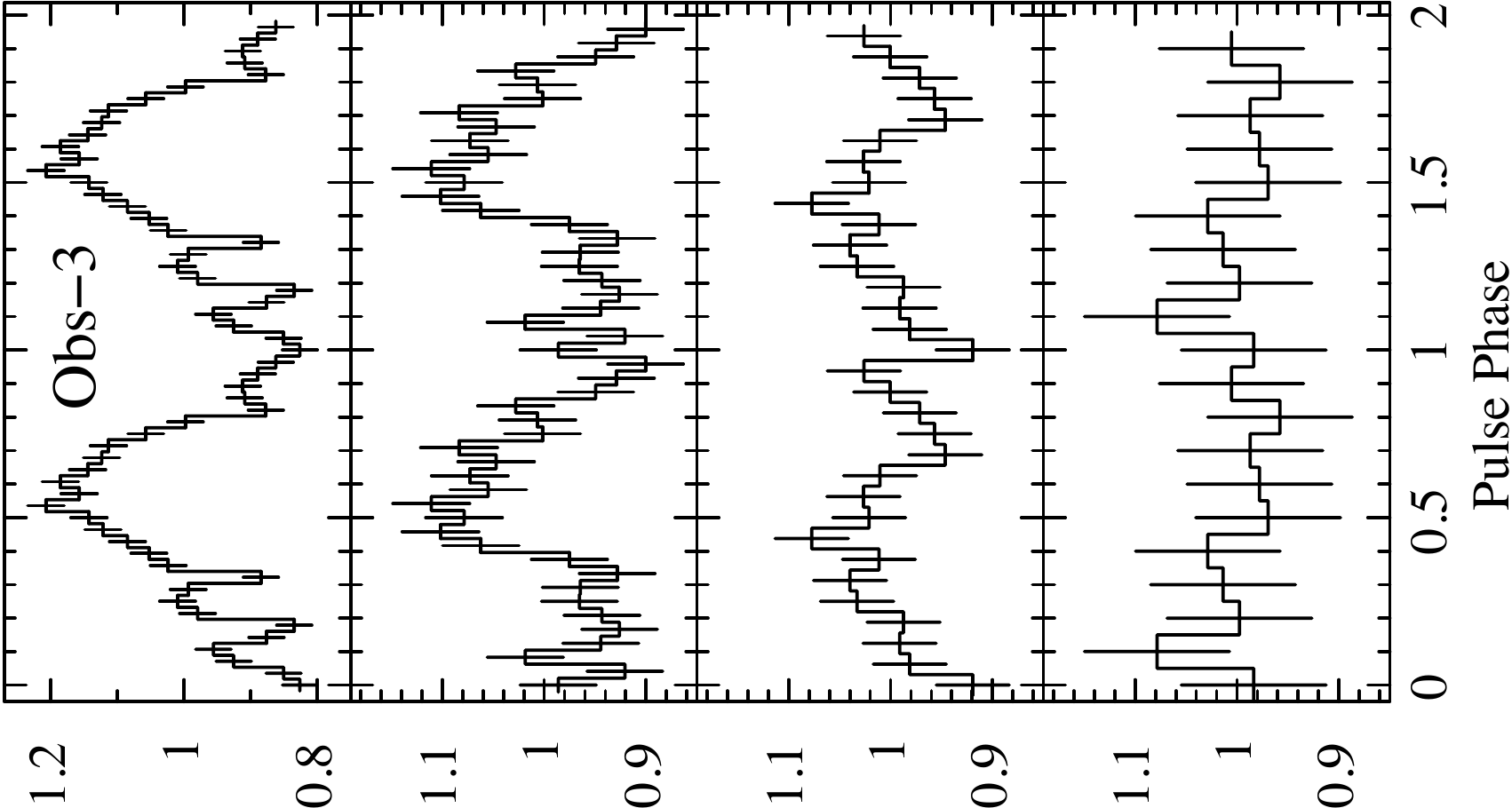}\\
\medskip
\includegraphics[width=.5\textwidth, angle=-90]{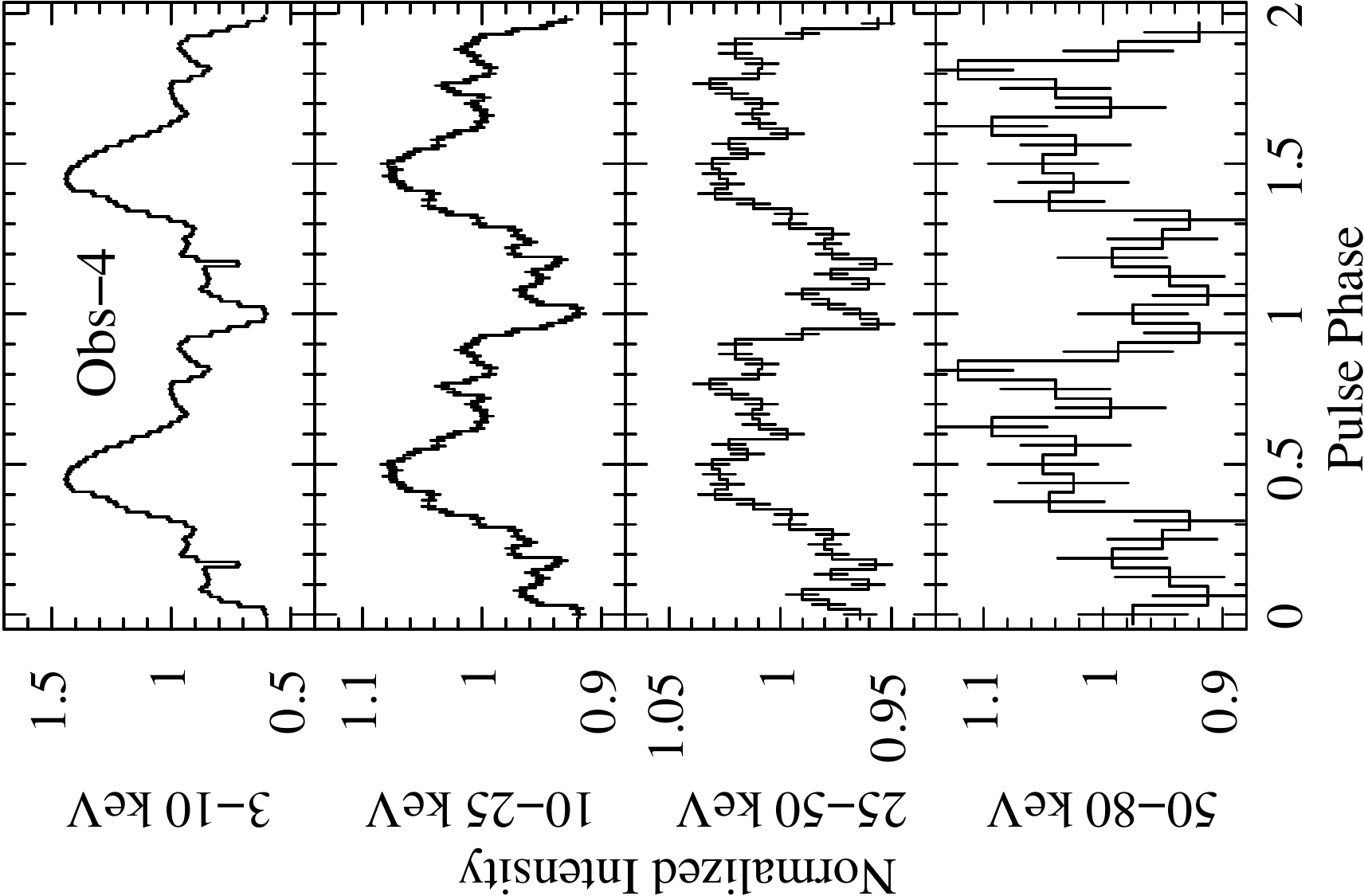}\quad
\includegraphics[width=.5\textwidth, angle=-90]{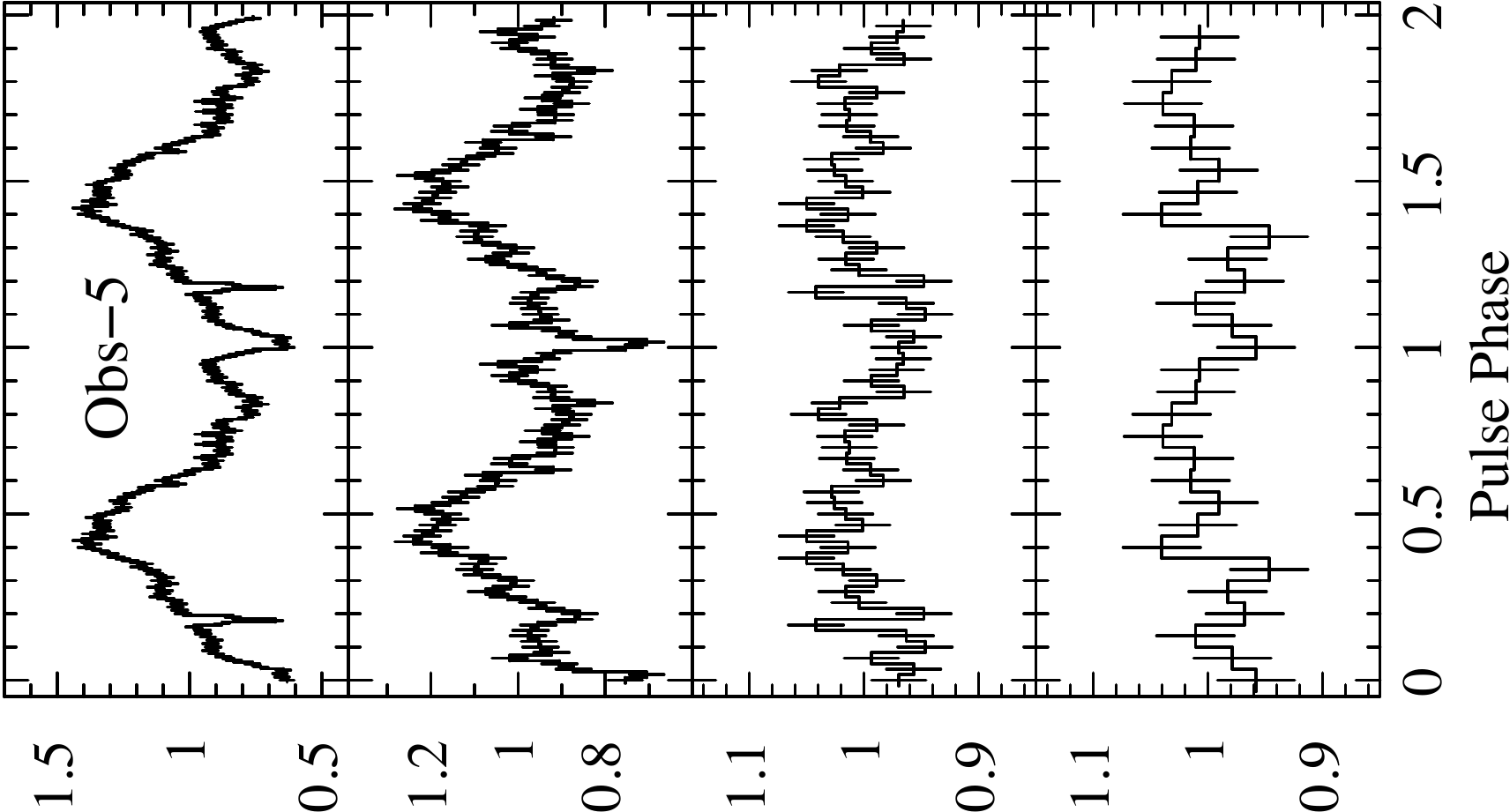}
\caption{Energy resolved pulse profiles of \source obtained by folding the energy resolved light curves from LAXPC instrument(s) onboard \astrosat at the respective estimated spin period(s). Two pulses are shown in each panel for clarity. The error-bars in the figure represent 1$\sigma$ uncertainties.}
\label{resolved-profiles}
\end{figure*}


\begin{figure}[t!]
\centering
\includegraphics[height=3.3in, width=2.5in, angle=-90]{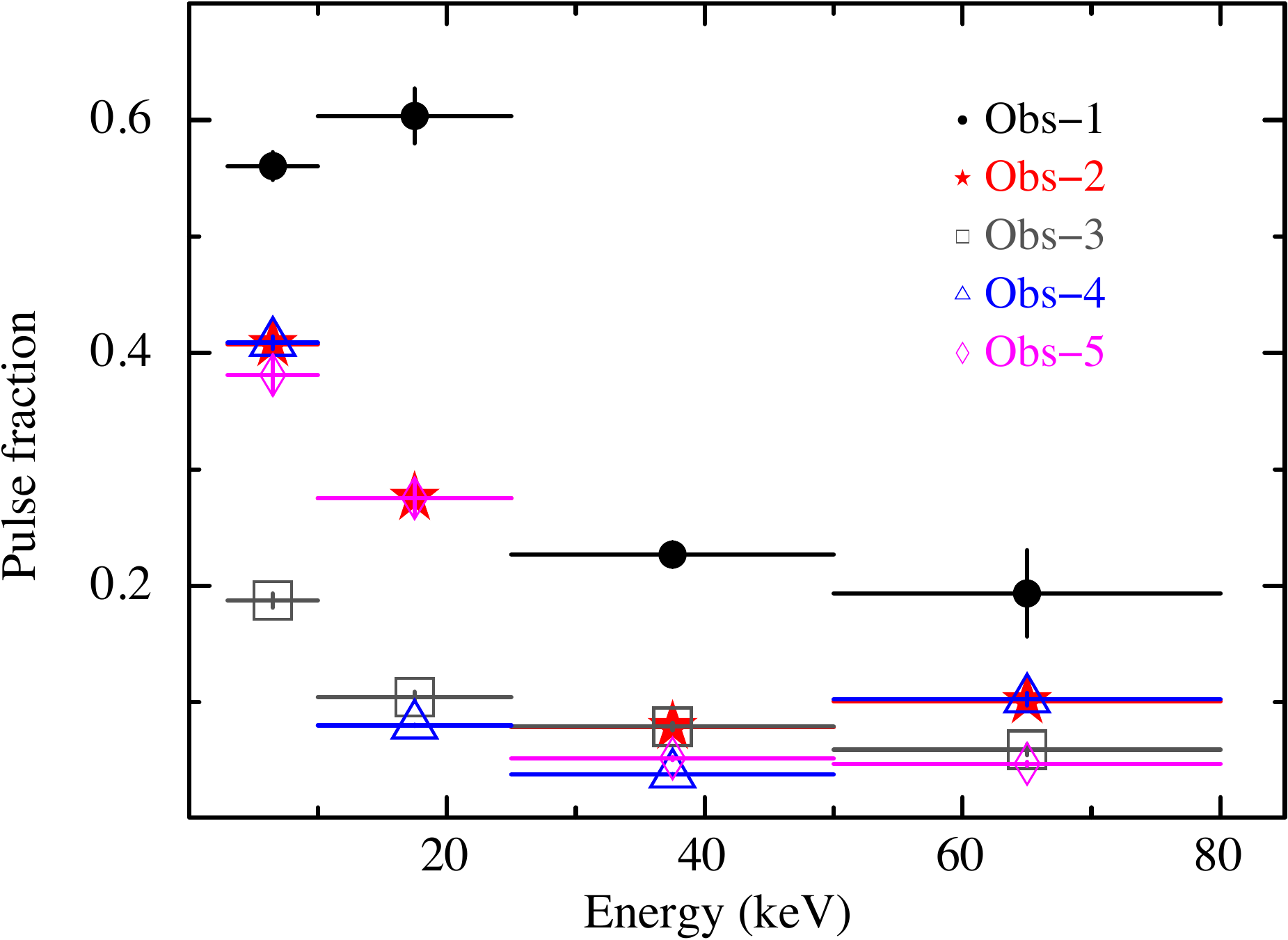}
\caption{Pulse fraction variation of \source with energy, obtained from the pulse profiles 
in multiple energy bands from five \astrosat observations.}
\label{pf}
\end{figure}

\begin{figure*}[bt!]
 \begin{center}$
 \begin{array}{cccccc}
 \includegraphics[width=0.35\textwidth, angle=-90]{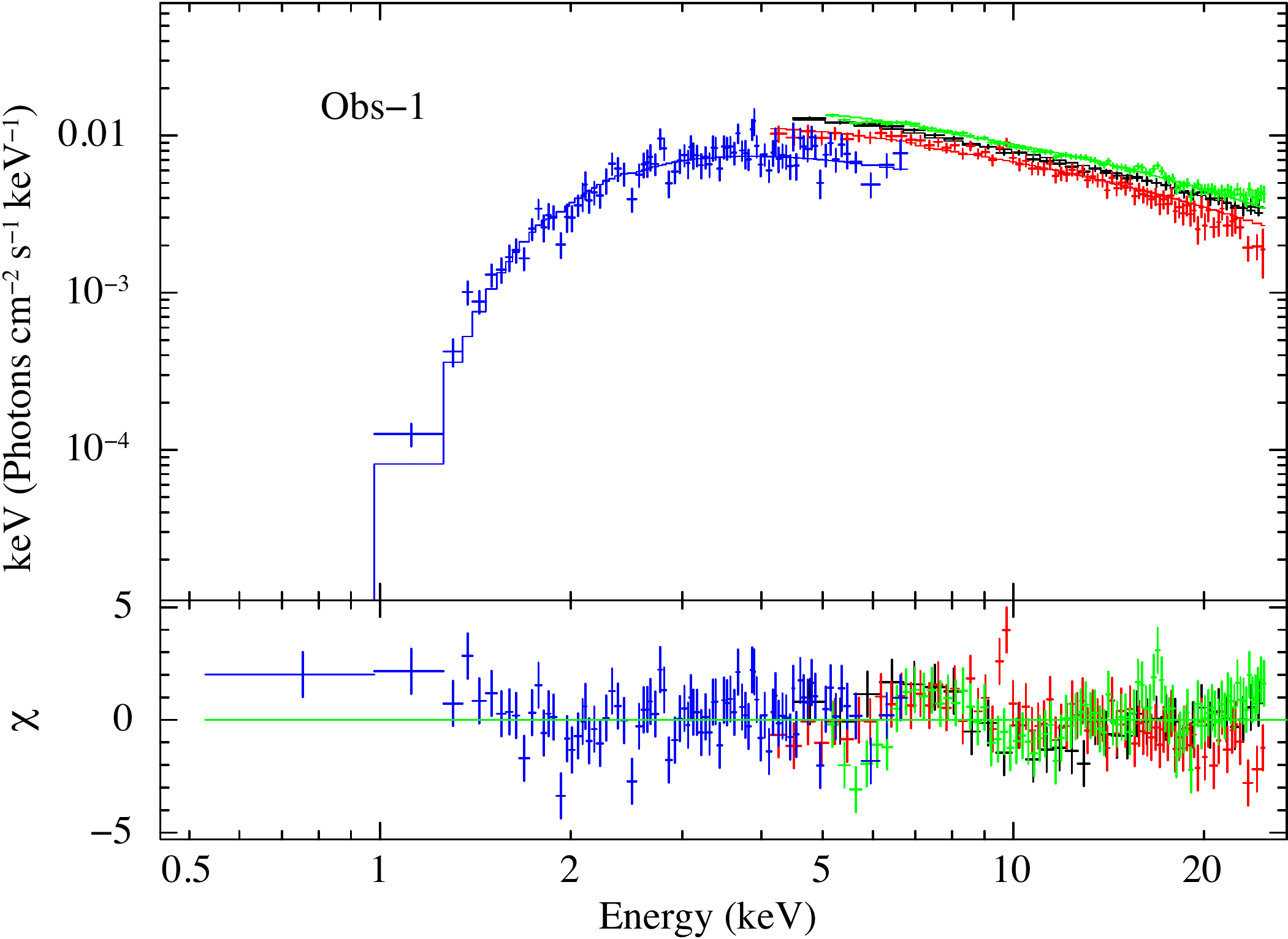}  &
 \includegraphics[width=0.35\textwidth, angle=-90]{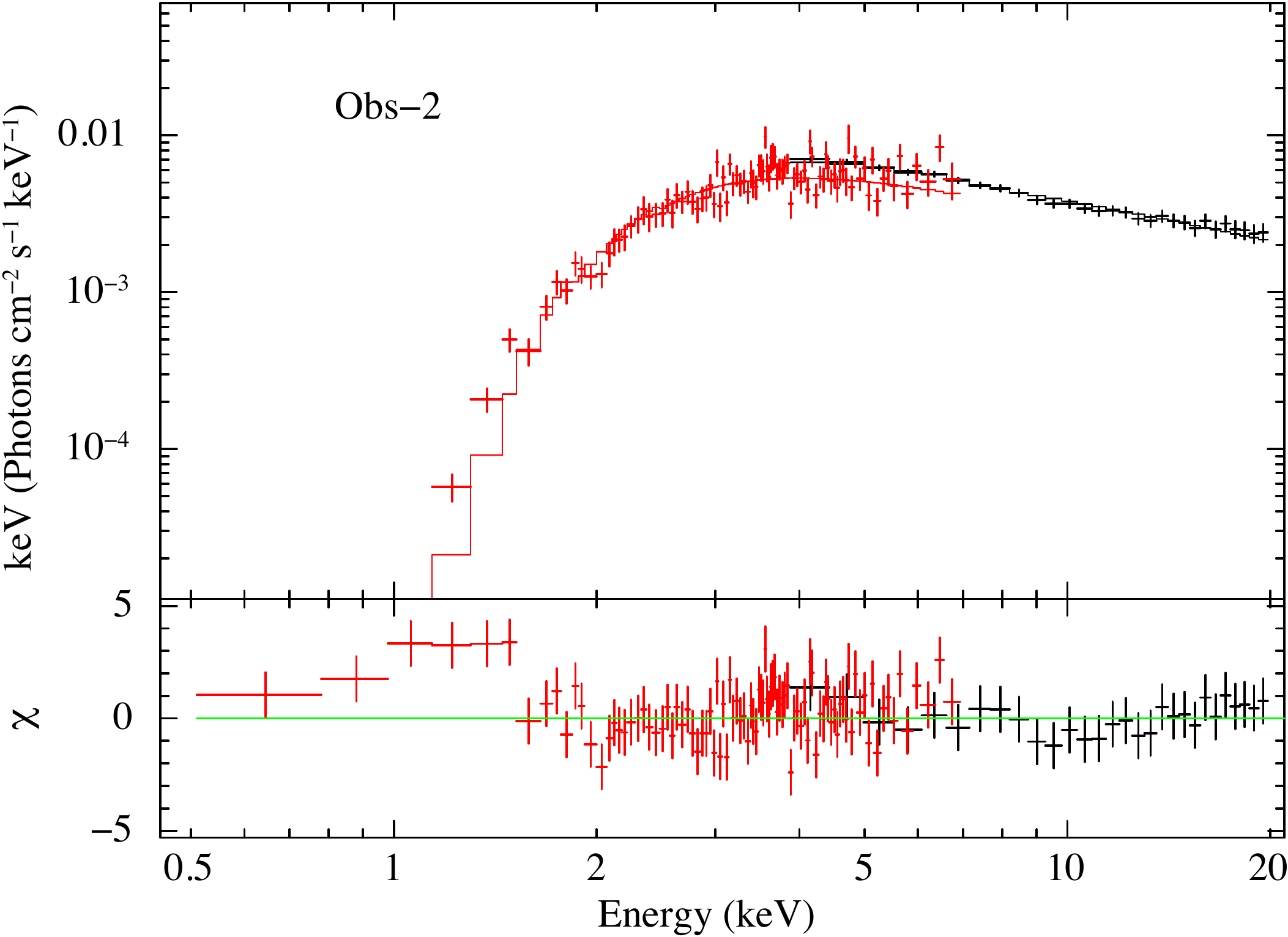} \\
  \includegraphics[width=0.35\textwidth, angle=-90]{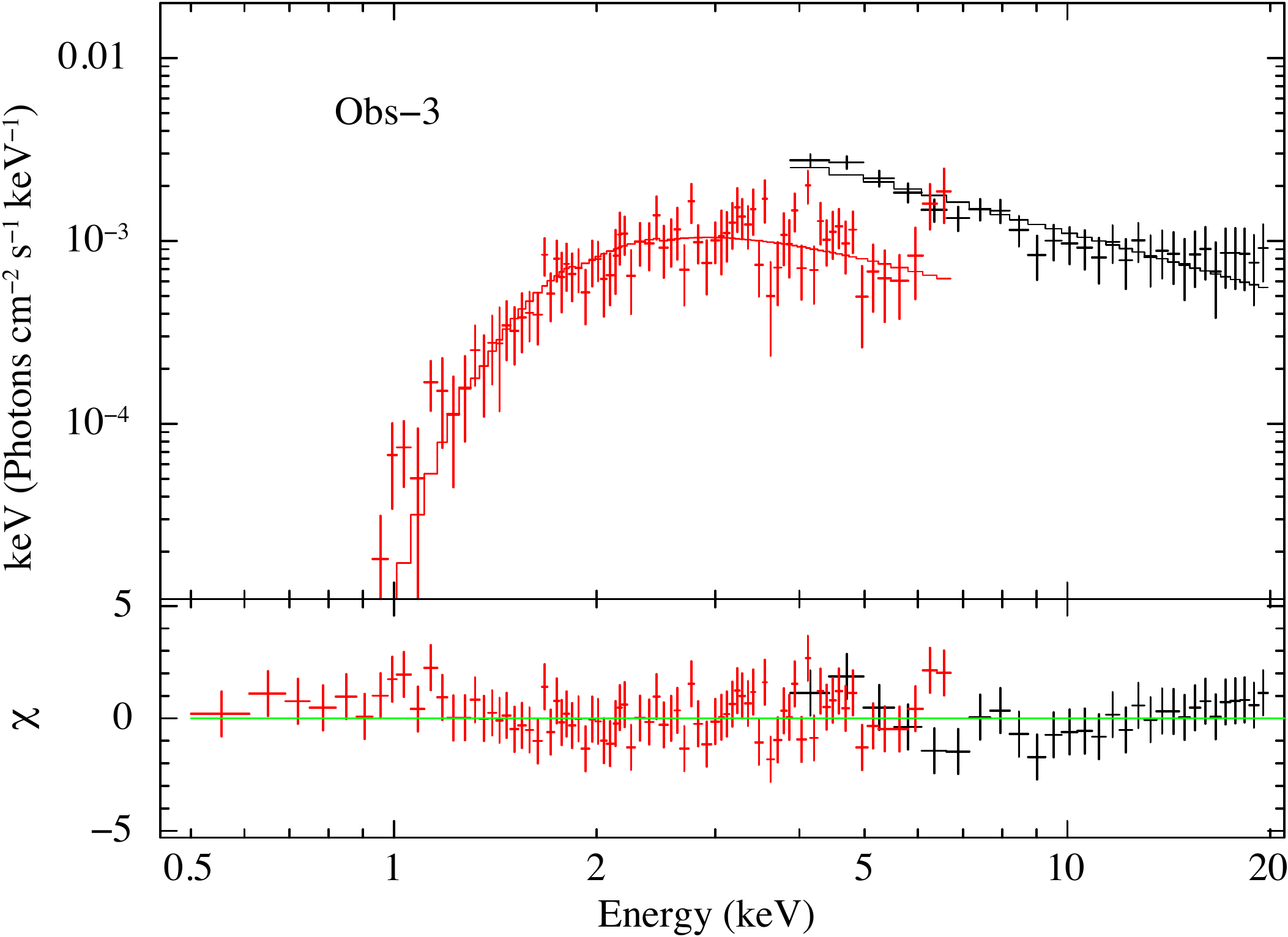} &
  \includegraphics[width=0.35\textwidth, angle=-90]{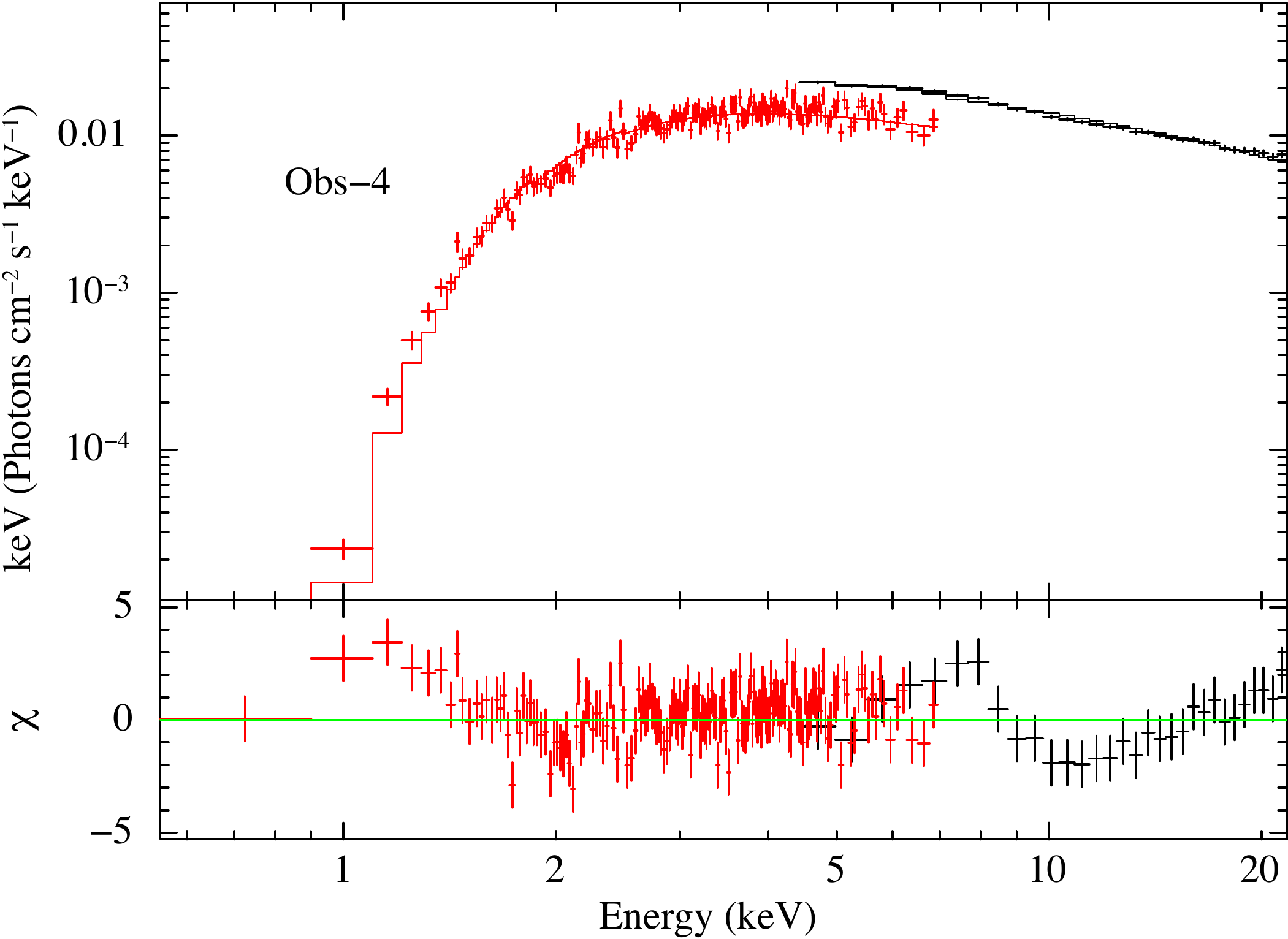} \\
    \includegraphics[width=0.35\textwidth, angle=-90]{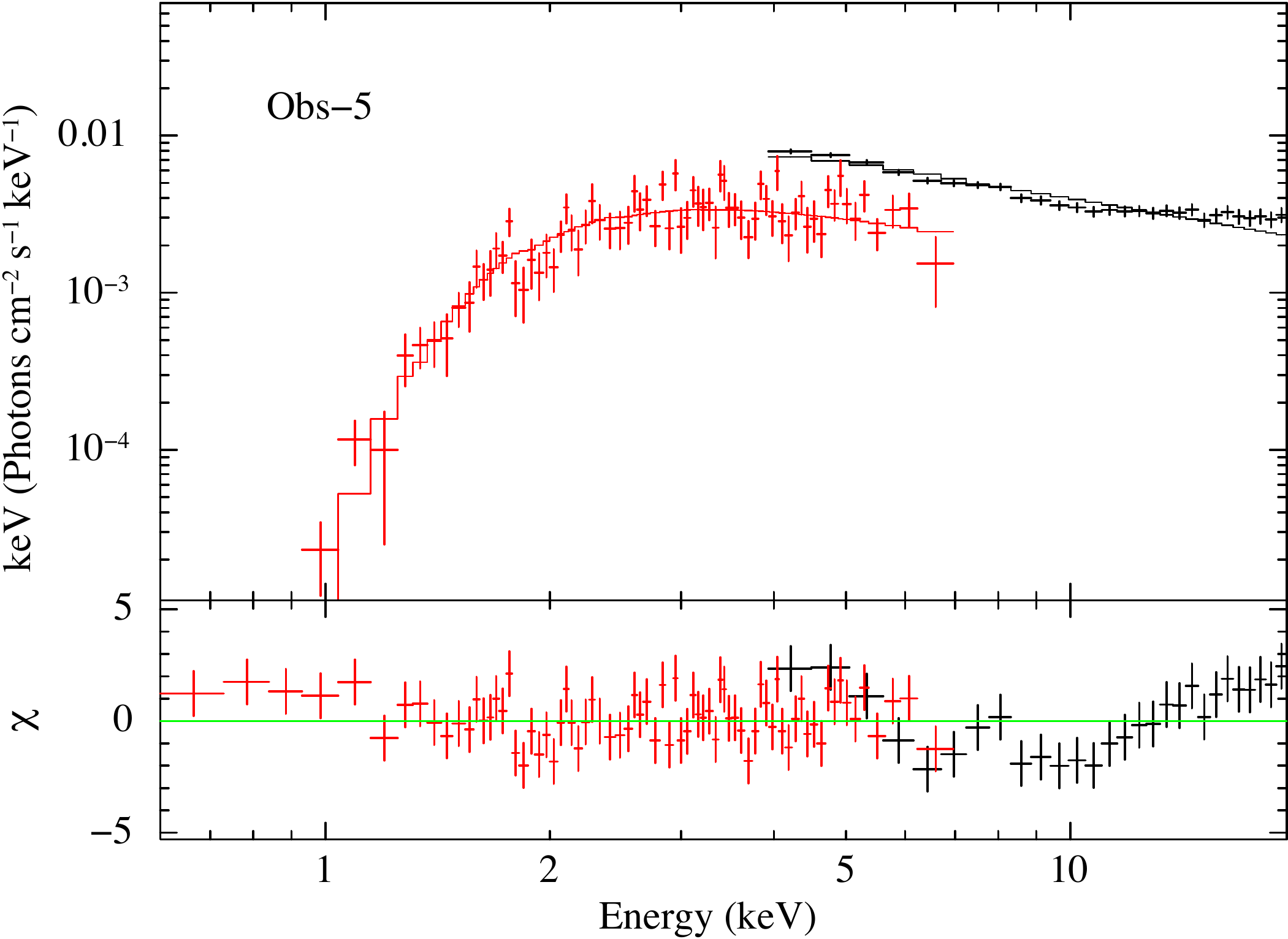} &
 \includegraphics[width=0.35\textwidth, angle=-90]{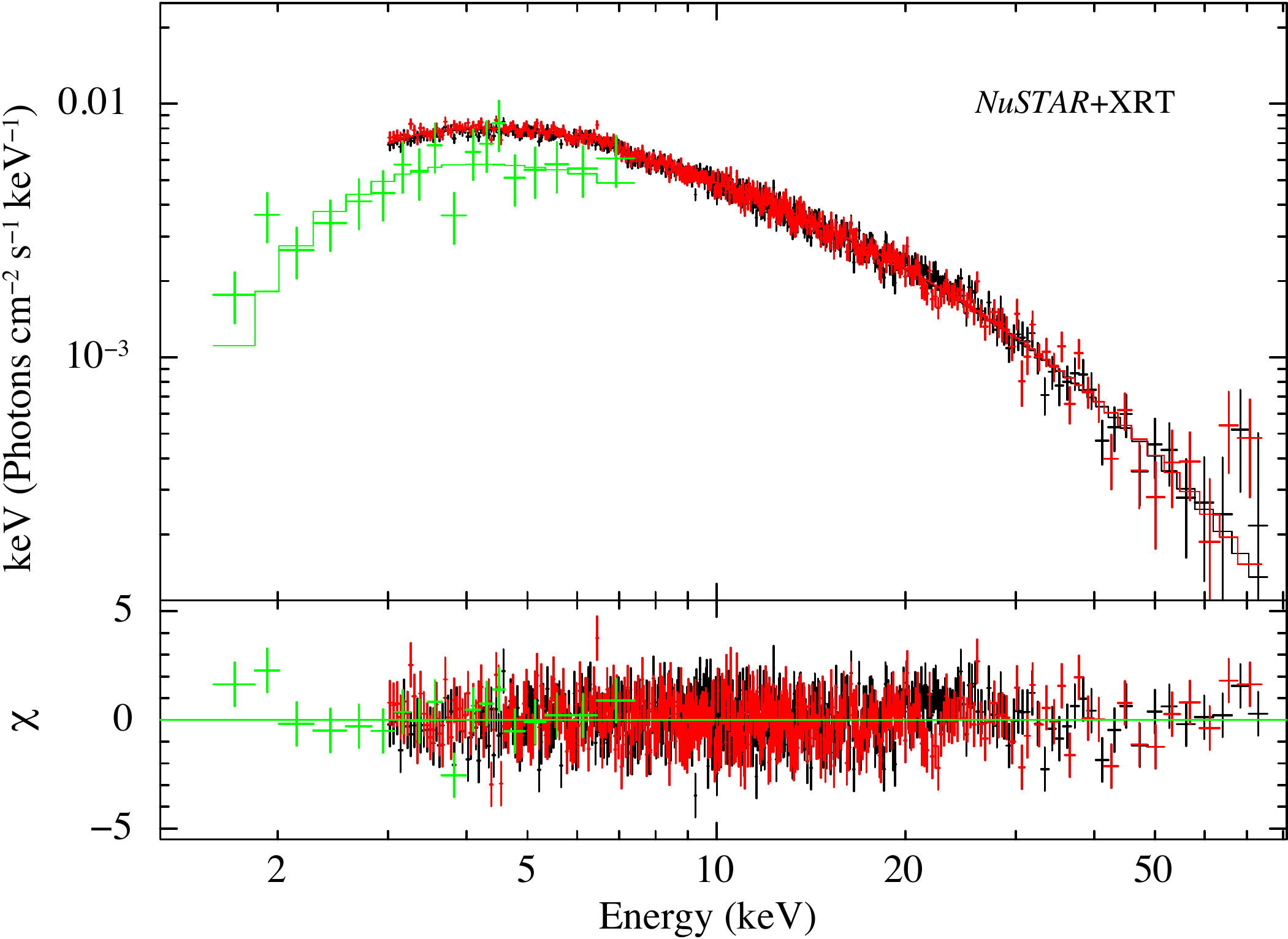} \\
 \end{array}$
 \end{center}
\caption{Best-fitting energy spectra obtained from first, second, third, fourth and fifth \astrosat observations of \source.  Broadband energy spectra from \nustar and {\em Swift}/XRT data in 2016 July are also shown.}
\label{spec}
\end{figure*}

\section{Observations and Data Reduction}

\subsection{\astrosat}

The first Indian multi-wavelength astronomical satellite, \astrosat was launched by Indian Space Research Organization on 28 September 2015 (Agrawal 2006, Singh et al. 2014). The observatory is sensitive to  photons from optical to hard X-ray ranges by using five sets of instruments such as Ultraviolet Imaging Telescope (UVIT; Tandon et al.  2017), Soft X-ray Telescope (SXT; Singh et al. 2017), Large Area X-ray Proportional Counters (LAXPCs; Agrawal et al. 2017, Antia et al.  2017), Cadmium Zinc Telluride Imager (CZTI; Rao et al.  2017), and a Scanning Sky Monitor (SSM; Ramadevi et al.  2018). However, in the present study, five epochs of \astrosat observations of \source with SXT and LAXPC instruments are used. Details of the observations are summarised in Table 1. As the source was very faint during all five epochs, it was not detected in the CZTI. The UVIT was not operational during these observations. \astrosat caught the source at different phases of the regular Type~I X-ray outbursts. Top and bottom panels of Figure~\ref{maxi-bat} show the MAXI (Monitor of All-sky X-ray Image, Matsuoka et al. 2009) and \swift/BAT (Burst Alert Telescope, Krimm et al. 2013) monitoring light curves of the pulsar covering the epochs of \astrosat observations. The first, fourth and fifth \astrosat observations were carried out at the declining phase of the Type-I X-ray outbursts at a source luminosity of 15-40 mCrab with BAT. However, during the second observation, monitoring data from MAXI or {\it Swift}/BAT were not available. An extremely low level of X-ray intensity, $\sim$10 mCrab in 15-50 keV range, was estimated during the third \astrosat observation.  A log of these \astrosat pointings of \source is given in Table~1. 

The SXT is a soft X-ray focusing telescope onboard \astrosat. It consists of shells of conical mirrors that focus the soft X-ray photons in 0.3--8~keV energy range on a CCD detector. The field of view of the SXT is 40 arcmin. The effective area of the telescope is 90~cm$^2$  at 1.5 keV. The energy resolution of the detector is 90 eV at 1.5 keV and 136~eV at 5.9~keV. The source was observed with SXT in the photon counting mode, yielding a time resolution of 2.4~s. We followed standard analysis procedure for the SXT data reduction as suggested by the \astrosat Science Support Cell (ASSC\footnote{\url{http://astrosat-ssc.iucaa.in/}}). The source spectrum was extracted from a 8~arcmin circular region centered at the source coordinate on the SXT chip using {\tt XSELECT} package. The background spectrum was extracted from the blank sky region on the chip.

The LAXPC is a proportional counter detector sensitive to X-ray photons in the 3--80 keV energy range. There are three identical detector units onboard \astrosat with an effective area of about 8000 cm$^2$ at 15~keV. The time and energy resolutions of these units are 10~$\mu$s and 12\% at 22~keV, respectively. Standard data analysis routines ({\tt LAXPCsoftware}) are used to obtain the source light curves and spectral products from the event mode data. We have used SXT and LAXPC data in our timing study. Depending on the quality of the LAXPC data and instrument gain stability, we have considered events from single or combined LAXPC units. For timing studies, combined data from LAXPC-10, 20 \& 30 are used during Obs-1, while data from LAXPC-20 only are considered for Obs-2, Obs-3, and Obs-5. The events from LAXPC-10 \& 20 are used for timing studies from Obs-4. Background products corresponding to each observation are accumulated from the same data by analysing the Earth occultation period. A systematic uncertainty of 2\% is also added in the LAXPC spectra. 

\subsection{\nustar and \swift/XRT}
In the present study, we also used \nustar (Harrison et al. 2013) and \swift/XRT (X-Ray Telescope; Burrows et al. 2005) observations on 25 July 2016, at a reported lowest luminosity of \source till date (F{\"u}rst  et al. 2017), to compare the results obtained from the \astrosat observations. For \nustar observation, we used the {\tt NuSTARDAS} 1.6.0 software in {\tt HEASoft} version 6.24. Unfiltered events from the FPMA and FPMB were reprocessed by using the {\it nupipeline} routine in the presence of CALDB of version 20191219. Source products were then extracted by selecting circular region of 120~arcsec radius with souce coordinates as center by using the {\it nuproducts} task. Background products were also accumulated in a similar manner by selecting a source-free region. Data from the \swift/XRT observation in photon counting mode, with an effective exposure of 1 ks, are also used. We obtained XRT products by using the online standard tool provided by the UK Swift Science Data Centre\footnote{\url{http://www.swift.ac.uk/user_objects/}} (Evans et al. 2009). 

\section{Timing Analysis}

We extracted source and background light curves from the SXT and LAXPC event data at 2.4~s and 0.1~s binning time, respectively. After subtracting the background, X-ray pulsations were searched in the barycentric corrected light curves of \source from all five observations. We applied the chi-square maximization technique using {\tt efsearch} task of {\tt FTOOLS} package (Leahy 1987). The spin period of the pulsar is estimated to be 41.2895(7) s, 41.272(9) s, 41.30(1) s,  41.2747(8) s, and  41.306(3) from first, second, third, fourth, and fifth \astrosat/LAXPC observations, respectively.  The spin period and its error are also estimated by using the Lomb-Scargle and Clean techniques in the publicly available {\tt PERIOD} package (Currie et al. 2014). This package has been used for period estimation in several other binary X-ray pulsars e.g. 4U~2206+54 (Torrejón et al. 2018), 2S~1417-624 (Gupta et al. 2019), Swift~J0243.6+6124 (Beri et al. 2021). The results obtained from these methods are found to agree with the above quoted values.  The evolution of pulse period with luminosity using {\it Astrosat} observations is presented in Figure~\ref{spin-period}. As the source was observed at a low luminosity level, a few measurements have large errors on the spin period. A marginal spinning-up with increasing luminosity can be seen in the figure, though it is not adequate to draw any significant claim on this.

The light curves in 0.3-7 keV and 3-80 keV ranges from the SXT and LAXPC data from each epoch of observations are folded with the corresponding estimated pulse period to obtain the pulse profiles of the pulsar. The pulse profiles obtained from the SXT and LAXPC data for all five \astrosat observations are shown in  Figure~\ref{sxt-profile} \& ~\ref{lxp-profile}, respectively. Phases of the pulse profiles are adjusted manually to align the minima at phase zero. The profiles obtained from the SXT data (Figure~\ref{sxt-profile}) appears single peaked. This is possibly due to the the fact that the soft X-ray photons are largely affected by absorption due to the material along the line of sight and low source count rate in SXT. The profiles from the LAXPC data, however, are found to be complex due to the presence of multiple structures at various pulse phases of the pulsar during the first, third, fourth and fifth observations (see Figure~\ref{lxp-profile}). Sharp dip-like features were detected in 0.1--0.2 and 0.60-0.85 phase ranges during these observations. Pulse profile of the pulsar from the second epoch of observations, however, appears relatively simpler. 

To investigate the observed features in the LAXPC profiles with energy, barycenter corrected light curves in 3-10, 10-25, 25-50, and 50-80~keV ranges are extracted from the LAXPC data from all epochs of observations and folded with the respective spin period and shown in Figure~\ref{resolved-profiles}. The energy resolved pulse profile are found to be strongly energy dependent. The presence of dip-like features are seen up to higher energies in the profiles from all observations. The observed dips are evident up to $\sim$50~keV, especially during the first observation, whereas during second, fourth and fifth observations, the features are present up to $\sim$25~keV. However, during the third observation, the dips are present up to $\sim$10 keV (Figure~\ref{resolved-profiles}).   We checked the significance of pulsations in the hard X-ray band by taking a ratio between the peak count rate and the standard deviation of the low or minimum intensity interval observed in the pulse profile. It is found that the significance of detection of pulsation in 50-80 keV range is more than 15$\sigma$ during all observations except the third observation.

We calculated the pulse fraction of the pulsar using the pulse profiles in various energy bands and presented in Figure~\ref{pf}. It is done to determine the nature of pulsating component. In our study, we define the pulse fraction as the ratio between the difference and sum of maximum and minimum intensities observed in the pulse profile of the pulsar. For all the observations, we found that the pulse fraction decreases with energy. A maximum value of pulse fraction of $\sim$60\% is detected in the profiles below 20 keV during the first observation. A relatively lower value is observed in rest of the data sets.


\begin{table*}
\tabularfont
\centering
\caption{Best-fitting spectral parameters (with 90\% errors) of \source during \astrosat and {\em NuSTAR}+XRT observations.}
\begin{tabular}{lccccccc}
\hline
\hline

Parameters           &Obs-1       &Obs-2     &Obs-3      &Obs-4    &Obs-5 &{\em NuSTAR}+XRT\\
	
\hline
N$_{\rm H}$ (10$^{22}$~cm$^{-2}$)        &4.5$\pm$0.3      &4.7$\pm$0.3      &3.5$\pm$0.5    &4.6$\pm$0.3    &4$\pm$0.4   &5.8$\pm$0.6\\
Photon index ($\Gamma$)                &1.61$\pm$0.07    &1.92$\pm$0.05  &2.1$\pm$0.2    &1.54$\pm$0.1   &1.85$\pm$0.05   &1.64$\pm$0.05  \\
Norm (10$^{-2}$)           &3.7$\pm$0.4      &3.4$\pm$0.3      &1.3$\pm$0.5    &6.5$\pm$1   &2.9$\pm$0.3   &2.5$\pm$0.2 \\
E$_{\rm fold}$ (keV)    &7.2$\pm$0.7   &--               &--                &6$\pm$2   &--               &6.5$\pm$0.4\\
E$_{\rm cut}$ (keV)	    &36$^{+8}_{-6}$   &--               &--             &32$^{+11}_{-8}$   &--        &27$\pm$2 \\
Flux$^a$ (3-30 keV)  	          &2.8$\pm$0.1      &1.5$\pm$0.1    &0.42$\pm$0.03  &5$\pm$0.1    &1.51$\pm$0.1     &1.67$\pm$0.1\\
Flux$^a$ (0.5-30 keV)  	          &3.1$\pm$0.1      &1.7$\pm$0.1    &0.50$\pm$0.05  &5.5$\pm$0.1   &1.71$\pm$0.1    &2$\pm$0.1\\
Luminosity$^b$ (10$^{36}$~\lumcgs)   	            &1.9              &1.0            &0.25           &3.3    &1.0                    &1.21\\
$\chi^2_\nu$ ($\nu$)  &1.06 (363)        &1.28 (258)     &0.91 (105)     &1.18 (454)  &1.48 (100)   &0.96 (808)   \\
\hline
\hline
\end{tabular}
\label{table-spec}
\tablenotes{$^a$: unabsorbed flux in 10$^{-10}$~\fluxcgs; $^b$: 0.5-30 keV unabsorbed luminosity at a distance of 7.1 kpc.\\

Note: By fitting \nustar and XRT data, we estimated the unabsorbed flux of \source in 3-10 and 0.5-79 keV ranges to be 8.8$\times$10$^{-11}$ and 2.3$\times$10$^{-10}$~\fluxcgs, respectively. This is for comparison with the quoted values in F{\"u}rst et al. (2017).\\
}
\end{table*}

\begin{figure}[t!]
\centering
\includegraphics[height=3.3in, width=2.7in, angle=-90]{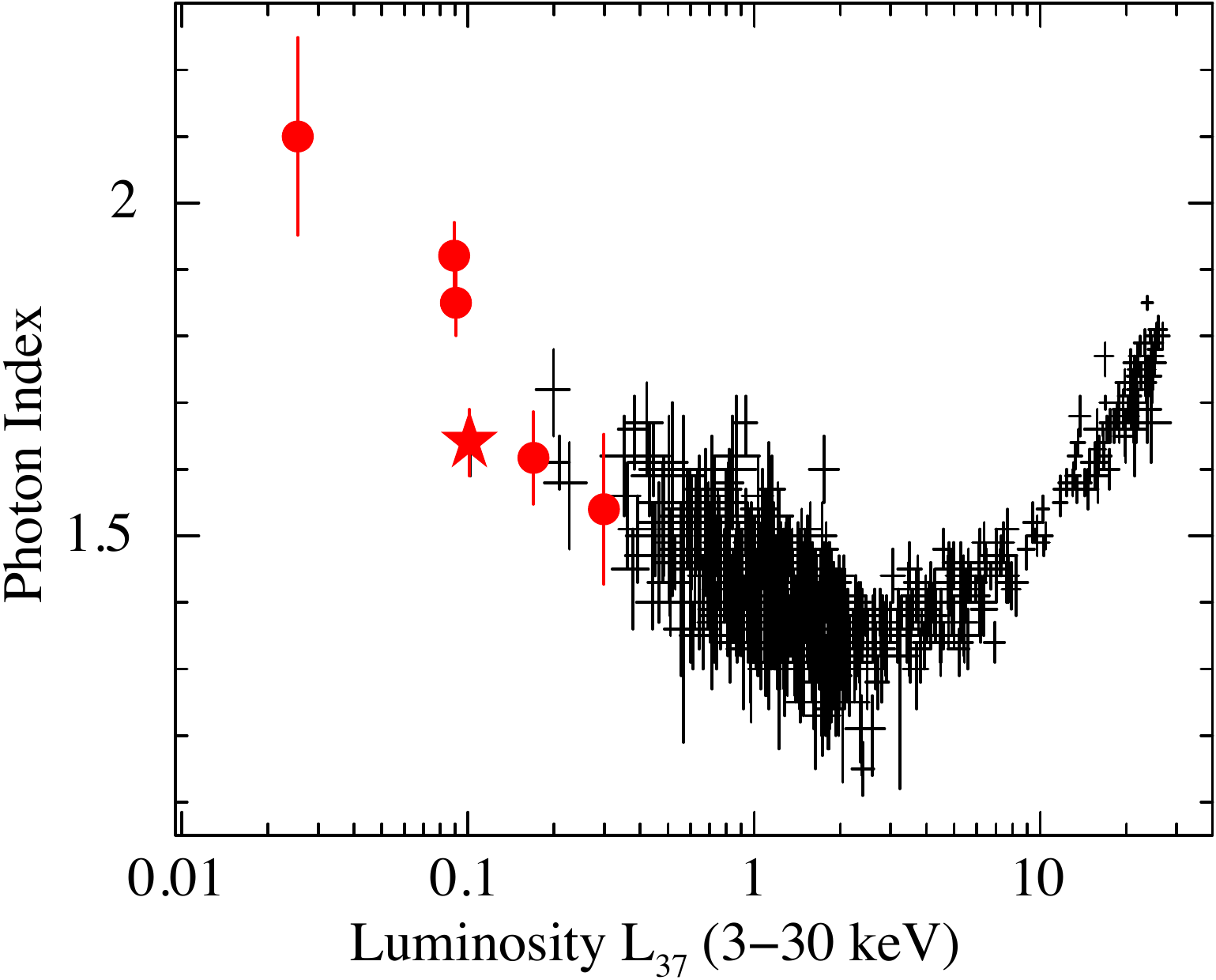}
\caption{Variation of power-law photon index with 3-30 keV luminosity are shown for {\em AstroSat} and {\em NuSTAR} observations of EXO~2030+375 with solid bullets and star symbols (red points), respectively, along with the corresponding data from the {\it RXTE} observations (black points) as shown in Figure~6 of Epili et al. (2017). The power-law photon index obtained from the present study follows the anti-correlated pattern with the luminosity in the sub-critical regime of the pulsar. L$_{\rm 37}$ denotes the 3-30~keV  unabsorbed luminosity of the pulsar in 10$^{\rm 37}$~\lumcgs~ at a distance of 7.1~kpc.}
\label{lum-ind}
\end{figure}

\section{Spectral Studies}

The spectral properties of \source are studied using data from all five \astrosat observations. Using the source and background spectra extracted from the SXT and LAXPC data (as described above) and response files provided by the instrument team, we carried out spectral fitting of the data in 0.5--7 keV from SXT and 3.5-25 keV from LAXPC by using {\tt XSPEC} package  of version 12.10.0 (Arnaud 1996). The LAXPC data were limited to 25 keV in our spectral fitting because of background uncertainties at high energies. Various standard models such as power-law, cutoff power-law, high energy cutoff power-law are attempted to fit the 0.5-25 keV spectrum, along with a component for photo-electric absorption ({\tt TBabs}, Wilms, Allen \& McCray 2000). We found that a cutoff based model is necessary to describe the spectra obtained from the first and fourth observations when the pulsar was relatively brighter  (Figure~\ref{maxi-bat}). However, a simple absorbed power-law model can describe the spectra from second, third and fifth observations, satisfactorily. These models provided goodness of fit per degree of freedom close to $\chi^2_\nu$=$\chi^2/\nu$ $\approx$1 in all cases. In place of a power-law component, we also tried to fit the spectra from second, third and fifth observations with a thermal blackbody component. This yielded a poor fit with a $\chi^2/\nu$ of more than 5. We do not detect any signature of cyclotron absorption scattering feature(s) (Jaisawal \& Naik 2017; Staubert et al. 2019) in 0.5-25 keV spectral range. The spectra obtained from SXT and LAXPC data also do not show any iron emission line(s) in 6-7 keV range. Spectral parameters estimated from these fittings are given in Table~2. In our fitting, the relative instrument normalization for SXT was found to be in the range of 0.65-0.80 with respect to LAXPC. 
 
We fitted the \nustar data from FPMA and FPMB detectors along with the \swift/XRT data in 1-79 keV energy range with a high energy cutoff power-law model along with the interstellar absorption. This model fitted the spectrum well. The spectral parameters such as column density, power-law photon index, cutoff and folding energies obtained from our fitting are found to be consistent with the values reported in Table~1 of F{\"u}rst  et al. (2017).      

The energy spectra corresponding to each observation along with best-fit model and corresponding residuals are shown in Figure~\ref{spec}. The {\tt cflux} convolution model is used for flux estimation in our study. We would like to mention here that the quoted flux and luminosity in Table~2 are estimated in 0.5-30 and 3-30 keV range though 0.5-25 keV data were used in spectral fitting with \astrosat. Estimation of flux and luminosity in 3-30 keV range was done for comparison with the earlier values reported in literature. It is done by extrapolating the best-fit model up to 30 keV.  We attempted to find any correlation between the power-law photon index and luminosity of the pulsar during the \astrosat observations. For this, we plotted the photon index with the observed 3--30 keV luminosity of the pulsar from \astrosat and combined \nustar and \swift/XRT observations in Figure~\ref{lum-ind}. Along with this, corresponding data from the \rxte observations of EXO~2030+375, as reported in the top left panel of Figure~6 of Epili et al. (2017), are also shown for comparison. The figure shows that the photon index and luminosity are anti-correlated during all five epochs of \astrosat and combined \nustar and XRT observations. The results obtained from the present work extended the observed spectral behaviour of \source in sub-critical regime at much lower luminosity.

\section{Discussion}

Be/X-ray binary pulsars are expected to show X-ray enhancements, termed as Type-I X-ray outbursts, at the periastron passage of the neutron star (Okazaki \& Negueruela 2001, Reig 2011). However, in most of the cases, such enhancements are not always observed which has been interpreted as due to the lack of significant evolution of the equatorial circumstellar disk around the Be star companion. An alternative interpretation of the lack of X-ray activities (Type-I or Type-II outbursts) is related to the Be-disk dynamics due to Kozai-Lidov effect (Laplace et al. 2017). \source is unique in the sense that the pulsar shows Type-I X-ray outburst almost at every periastron passage of binary. The pulsar had been observed and studied extensively with the {\it Rossi X-ray Timing Explorer (RXTE)} for 606 epochs, spanning over 15 years, during Type-I and Type-II (giant) X-ray outbursts (Epili et al. 2017) though there are many pointing observations with other observatories used to study the characteristics of the source. Long term monitoring data from {\it RXTE}/ASM, {\it Swift}/BAT and MAXI/GSC show regular Type~I X-ray outbursts at the periastron passage of the pulsar. However, it has been noticed that the intensity at the peak of the Type-I X-ray outbursts has been in decline since last several years (Naik \& Jaisawal 2015, F{\"u}rst et al. 2016, Laplace et al. 2017) along with an extended period of low X-ray activity without any Type-I outbursts from 57200 MJD (27 June 2015) to 57600 MJD (31 July 2016)  (Kretschmar et al. 2016). Following the extended low state, the transient activities started with the appearances of outbursts, though of lower peak intensities to date.

\subsection{Detection of X-ray pulsations at the lowest observed luminosity}

The {\it NuSTAR} and \swift/XRT observations on 25 July 2016 were reported to be carried out at the lowest luminosity of \source during which pulsations were detected in the light curves (F{\"u}rst et al. 2017). Though the pulsar was observed at even lower luminosity of $\sim$10$^{34}$ erg s$^{-1}$ with the {\it Swift}/XRT in 3--10 keV range, poor data quality refrained from pulsation search (F{\"u}rst et al. 2017). On reanalysis of the \nustar plus \swift/XRT data, the 0.5-30 keV range luminosity of the pulsar on 25 July 2016 was estimated to be 1.21$\times$10$^{36}$ erg s$^{-1}$ (Table~2). On comparing the luminosities during five \astrosat observations with the \nustar and \swift/XRT observations, it is interesting to point out that the pulsar was caught at even lower luminosities during second, third and fifth \astrosat observations. Among these, the lowest luminosity of 2.5$\times$10$^{35}$\lumcgs in 0.5-30 keV range was estimated during the third epoch of \astrosat observation. The LAXPC data from this observation showed a clear pulsation at 41.3~s in the light curve. Since the discovery in 1985, the luminosity of 2.5$\times$10$^{35}$\lumcgs in 0.5--30 keV range, observed with \astrosat/LAXPC on 19 June 2018 is the lowest luminosity of \source at which X-ray pulsations are detected in the light curves.

In accretion powered X-ray pulsars, material is channeled from the disk to magnetic poles. Decrease in the mass accretion rate decreases the ram pressure, eventually leading to the increase in the size of the magnetosphere (Illarionov \& Sunyaev 1975, Nagase et al. 1989). In case magnetosphere exceeds beyond co-rotation radius, the centrifugal barrier prohibits accreting material to fall onto the neutron star. This leads to the cessation of pulsations of the pulsar and is referred as propeller effect (Illarionov \& Sunyaev 1975).  Though, \source was detected at a lowest luminosity level of $\sim$10$^{34}$~erg~s$^{-1}$ using {\em Swift}/XRT data (F{\"u}rst et al. 2017), non-detection of X-ray pulsations and presence of softer thermal component with a temperature of 1.22~keV suggest the neutron star surface to be the source of observed emission, which occurs when the neutron star enters into the propeller phase (see, e.g., Wijnands \& Degenaar 2016, Tsygankov et al. 2016, F{\"u}rst et al. 2017). This allowed us to consider the luminosity of  2.5$\times$10$^{35}$\lumcgs (third \astrosat observation) as the lowest during which pulsations are seen. Assuming above luminosity as the upper limit for the onset of propeller effect, we can calculate the pulsar magnetic field as follows (Campana et al. 2002, F{\"u}rst et al. 2017) 
\begin{equation}\label{eq}
L_{lim}=7.3 \times k^{7/2} P^{-7/3}~R{_{6}^5}~B_{12}^{2}~M_{1.4}^{-2/3} \times 10^{37} {\rm erg~s^{-1}} 
\end{equation}   
where $P$ is spin period in $s$, $B_{12}$ is magnetic field in unit of 10$^{12}$~G, $R_6$ is the neutron star radius in unit of 10$^{6}$~cm, $M_{1.4}$ is the mass of the neutron star in unit of 1.4 \msol. The factor $k$ is related to the accretion geometry with a value of $k$=0.5 and 1 in case of disk and spherical wind accretions, respectively. Using above equation and assuming disk accretion scenario for EXO~2030+375, we obtain a range of equatorial magnetic field between (3--15)$\times$10$^{12}$~G for a minimum luminosity of $\approx$1$\times$10$^{34}$ and 2.5$\times$10$^{35}$\lumcgs, respectively. Based on the detection of cyclotron line, the polar magnetic field of the neutron star is tentatively estimated to be  1$\times$10$^{12}$~G (Wilson et al. 2008) and 5$\times$10$^{12}$~G (Klochkov et al. 2008). However,  later studies did not confirm the cyclotron feature in a broad  energy range (Naik et al. 2013, Naik \& Jaisawal 2015, F{\"u}rst et al. 2017). In absence of firm detection of cyclotron line, we calculate the magnetic field by putting standard parameters of a neutron star in Equation~\ref{eq} and found that \source hosts a highly magnetized neutron star with a field strength between (3--15)$\times$10$^{12}$~G.

\subsection{Pulse profiles and spectroscopy}

Extensive studies of \rxte observations of \source revealed that the pulse profiles of the pulsar strongly depend on the luminosity or mass accretion rate (Epili et al. 2017). Irrespective of the type of X-ray outbursts, whether regular Type-I or giant Type-II, or phases of the outbursts, the morphology of the pulse profiles remain same at a certain luminosity. In the present study, the pulse profiles of the pulsar during all five epochs of \astrosat observations are characterised with the presence of narrow dip-like features, though the features are prominent in the lowest luminosity phase on 19 June 2018. These features are commonly seen in Be/X-ray binary pulsars (see, e.g., Devasia et al. 2011, Jaisawal, Naik \& Epili 2016, Gupta et al. 2018, Jaisawal et al. 2018). At a luminosity an order of 10$^{36}$~\lumcgs, Ferrigno et al. (2016) and F{\"u}rst et al. (2017) have detected a sharp absorption feature in the pulse profile of EXO~2030+275. This feature is interpreted as due to the effect of obscuration through accretion column along the line of sight. This is supported by the phase-resolved spectroscopy which revealed a high column density and effectively harder spectrum due to reprocessing of the emission. In our study, the low luminosity of the pulsar and limited understanding of the background and spectral calibration of the instruments at high energies (Antia et al. 2017) prevented us to investigate the cause of the prominent dips in the pulse profiles through pulse-phase resolved spectroscopy. From energy resolved pulse profile (Figure~\ref{resolved-profiles}), clear pulsations up to $\sim$50 keV are seen during all five epochs of observations. Significance of pulsation can also be seen from the values of pulse fraction with energy (Figure~\ref{pf}). The fraction of number of photons contributing towards pulsation can be found to decrease with energy as well as luminosity. 

Broad-band energy spectrum of accretion powered X-ray pulsars originates due to thermal and bulk Comptonization of soft X-rays photons from the thermal mound on the neutron star surface (Becker \& Wolff 2007). In spite of complex processes taking place in the accretion column, the observed spectrum can be described by high energy cutoff power-law, exponential cutoff power-law models along with components for emission lines and absorption due to the interstellar medium. We have studied five \astrosat and {\em NuStar}+XRT observations between 2016 and 2020 after renewed activities from EXO~2030+375. Spectral analysis of these observations revealed the dependence of power-law photon index with luminosity. Extensive studies of available {\it RXTE} observations of the pulsar established the relation between the power-law photon index with source luminosity (Epili et al. 2017). From above study, the photon index are found to be distributed in three distinct regions depending on the 3-30 keV luminosity, suggesting the spectral transition from sub-critical to super-critical regimes through the critical luminosity of (2-4)$\times$10$^{37}$\lumcgs for \source at constant photon index. The source spectrum became harder with the luminosity in the sub-critical regime. A softening in the spectral emission was thereafter detected in the super-critical regime of the pulsar. As quoted, \astrosat observations were carried out at lower luminosities compared to the \rxte observations. In this study, we found that the power-law photon index is anti-correlated with the luminosity of the pulsar (Figure~\ref{lum-ind}) in the same manner as reported by Epili et al. (2017) at lower luminosities. This confirms that the the spectral shape of the pulsar depends on the mass accretion rate.

\section{Conclusion}
In this paper, we carried out timing and spectral studies of \source using five \astrosat observations at various phases of its Type-I X-ray outbursts. The source luminosity was detected to as low as  2.5$\times$10$^{35}$~\lumcgs in 0.5-30 keV range at which clear pulsations are detected. This is the first time when pulsations at such a low luminosity level is detected in the pulsar. Considering this as an limiting luminosity for propeller regime, we calculated the magnetic field of the neutron star.  We have also studied pulse profiles of the pulsar. The pulse morphology is found to be complex due to presence of multiple absorption like features. The energy spectrum of \source can be described by a high energy cutoff power-law model during brighter (first and fourth) \astrosat observations. The power-law photon index shows an anti-correlation with the source luminosity which is expected when the source is below the critical luminosity.

\vspace{2em}

\section*{Acknowledgements}

We thank the anonymous reviewer for suggestions on the paper. This publication uses the data from \astrosat mission of the ISRO, archived at the Indian Space Science Data Centre. We thank members of SXT and LAXPC instrument teams for their contribution to the development of the instruments and analysis software. The SXT and LAXPC Payload Operations Centers (POCs) at TIFR are acknowledged for verifying and releasing the data via the ISSDC data archive and providing the necessary software tools for data analyses. We also acknowledge the contributions of the \astrosat project team at ISAC and IUCAA. This research has made use of data obtained through HEASARC Online Service, provided by the NASA/GSFC, in support of NASA High Energy Astrophysics Programs. This work used the NuSTAR Data Analysis Software ({\tt NuSTARDAS}) jointly developed by the ASI Science Data Center (ASDC, Italy) and the California Institute of Technology (USA).
\appendix

\vspace{1em}


\end{document}